\documentclass[aps,prl,twocolumn,superscriptaddress,10pt]{revtex4-1}

\usepackage{graphicx}
\usepackage{mathtools}
\usepackage{amssymb,amsmath}
\usepackage[caption = false]{subfig}
\usepackage{dsfont}
\usepackage{booktabs}
\usepackage{placeins}
\usepackage{verbatim}
\usepackage{comment}
\usepackage{color}
\usepackage{float}
\usepackage{multirow}
\usepackage[utf8]{inputenc}
\usepackage[export]{adjustbox}

\newcommand{\reffig}[1]{Fig.~\ref{fig:#1}}

\newcommand{\refeqs}[1]{Eq.~(\ref{eq:#1})}

\newcommand{\uv}[1]{{\,\bf \hat{#1}}}
\newcommand{\rsym}[0]{\mathcal{R}}
\newcommand{\tribra}[1]{\langle{#1}\rangle}

\newcommand{\etal}[0]{\textit{et al.}\textcolor{white}{a}}

\newcommand{\Rey}[1]{${Re}$}

\newcommand{\PkD}[2]{UPO$_{{#1}{\textrm{#2}}}$}

\newcommand{\iden}[0]{\mathds{1}}
\newcommand{\inp}[0]{{\mathcal{I}}}
\newcommand{\dis}[0]{{\mathcal{D}}}
\definecolor{asparagus}{rgb}{0.53, 0.66, 0.42}
\definecolor{gray}{rgb}{0.5,0.5,0.5}

\begin{document}

\title{Capturing Turbulent  Dynamics and Statistics in Experiments \\ with Unstable Periodic Orbits\\}

\date{\today}

\author{Balachandra Suri}
\affiliation{IST-Austria, 3400 Klosterneuburg, Austria}
\author{Logan Kageorge}
\affiliation{School of Physics, Georgia Institute of Technology, Atlanta, Georgia 30332, USA}
\author{Roman O. Grigoriev}
\affiliation{School of Physics, Georgia Institute of Technology, Atlanta, Georgia 30332, USA}
\author{Michael F. Schatz}
\affiliation{School of Physics, Georgia Institute of Technology, Atlanta, Georgia 30332, USA}
\begin{abstract}

In laboratory studies and numerical simulations, we observe clear signatures of unstable time-periodic solutions in a moderately turbulent quasi-two-dimensional flow.  We validate the dynamical relevance of such solutions by demonstrating that turbulent flows in both experiment and numerics transiently display time-periodic dynamics when they {shadow}  unstable periodic orbits (UPOs). 
We show that UPOs we computed are also statistically  significant, with turbulent flows spending a sizable fraction of the total time near these solutions. 
As a result, the average rates of energy input and dissipation for the turbulent flow and frequently visited UPOs differ only by a few percent.
\end{abstract}

\maketitle

{Characteristic flow patterns (coherent structures) embedded in turbulence play critical roles in both moderately \cite{hof_2004} and highly turbulent flows \cite{robinson_1991, dennis_2014}, including cascade processes in  two and three dimensions \cite{boffetta_2012,vanVeen_2019, goto_2012}.  However, inherently statistical descriptions of turbulence, which are currently widely accepted, fail to describe coherent structures effectively. Consequently, they are unable to quantitatively predict  statistical averages of turbulent flows (e.g., energy dissipation rates).}

{Recent studies suggest that coherent structures in turbulence can be described by \textit{recurrent} (e.g., time-periodic) 
solutions to the deterministic equations governing fluid flow \cite{kawahara_2001,hof_2004,gibson_2008,lozar_2012,avila_2013,suri_2017a}. 
The existence of such solutions embedded within a chaotic set suggests the possibility of a fundamentally dynamical theory, inspired by Hopf's vision of turbulence as a walk between neighborhoods of recurrent solutions \cite{hopf_1948, gibson_2008}.
For  certain (e.g., uniformly hyperbolic) low-dimensional dynamical systems exhibiting chaos, this viewpoint has been fleshed-out; chaotic trajectories in state space {\it shadow} (follow)  {a dense set of recurrent} solutions in the form of unstable time-periodic orbits (UPOs).
This property enables short-time forecasting and the computation [via periodic orbit theory (POT)] of statistical averages from properly weighted sums evaluated over UPOs, with higher weights assigned to more frequently visited UPOs  \cite{auerbach_1987,cvitanovic_1988,lan_2010}.}

{Although the equations governing turbulence are formally infinite-dimensional, turbulent flows (due to dissipation ) can be represented as state space trajectories confined to finite-dimensional chaotic sets \cite{hopf_1948}.
This dimension can be estimated, e.g., based on the number of unstable directions of UPOs and can be relatively low $[O(10)]$ for transitional flows in domains of moderate size  \cite{channelflow,chandler_2013,budanur_2017a,suri_2019}. While this qualitative similarity with low-dimensional chaos is encouraging, {variability in the number of unstable directions for UPOs suggests turbulent flows} are nonhyperbolic \cite{kostelich1997}. The stable and unstable manifolds of dynamically-invariant sets become  tangent at some locations inside the chaotic set, destroying the shadowing property there and raising questions regarding the utility of UPOs for both forecasting and computing statistical averages.} 

To date, research devoted to developing and testing a dynamical description of turbulence based on UPOs has relied exclusively on direct numerical simulations (DNS) \cite{kawahara_2001,toh_2003,viswanath_2007, duguet_2008a,vanveen_2011b, kreilos_2012,avila_2013,willis_2013,budanur_2017a, vanVeen_2019,page_2020}. Despite the likely presence of nonhyperbolicity, studies focusing on transitional flows (with dynamics and statistics  dominated by coherent structures) have generated valuable new insight.
In canonical three-dimensional shear  flows (e.g., plane-Couette) it was shown that UPOs capture salient dynamical aspects (e.g., self-sustaining processes \cite{waleffe_1997}) and statistical averages (e.g., mean flow profile)  of turbulent flows \cite{kawahara_2001, toh_2003, viswanath_2007, avila_2013,vanVeen_2019}. 
However, definitive evidence in support of POT has not emerged even from studies  that identified large  sets of ($\approx50$) UPOs  \cite{chandler_2013,lucas_2015}.  

Previous numerical studies have imposed numerous flow restrictions, including spatially-periodic boundary conditions, minimal-flow-unit domains and symmetry-invariance, that are not representative of experiment. 
{Consequently, direct experimental evidence for shadowing--turbulent flows approaching UPOs and mimicking their \textit{spatiotemporal} evolution--has not been reported  previously. Also, some amount of noise is always present in experiments and how it affects the dynamical relevance of UPOs is not currently understood. Lastly, the statistical significance of  UPOs in laboratory flows is also an outstanding question.}

{In this Letter, we report clear evidence of UPOs in an experimental quasi-two-dimensional (Q2D) flow, in a domain whose size is much larger than a minimal flow unit.
DNS of this moderately turbulent (transitional) flow is performed with no-slip boundary conditions and without imposing any symmetry constraints.
In particular, to test the shadowing hypothesis,} we study the {spatiotemporal} evolution of turbulent flows that approach UPOs closely. 
We investigate the relation between statistical ``weights'' predicted by POT and how frequently turbulent flow approaches UPOs. Finally, we compare time-averaged properties of turbulent flows with those computed from  UPOs.

\begin{figure}[!b]
\centering
\subfloat[]{\includegraphics[width=1.3in,valign=c]{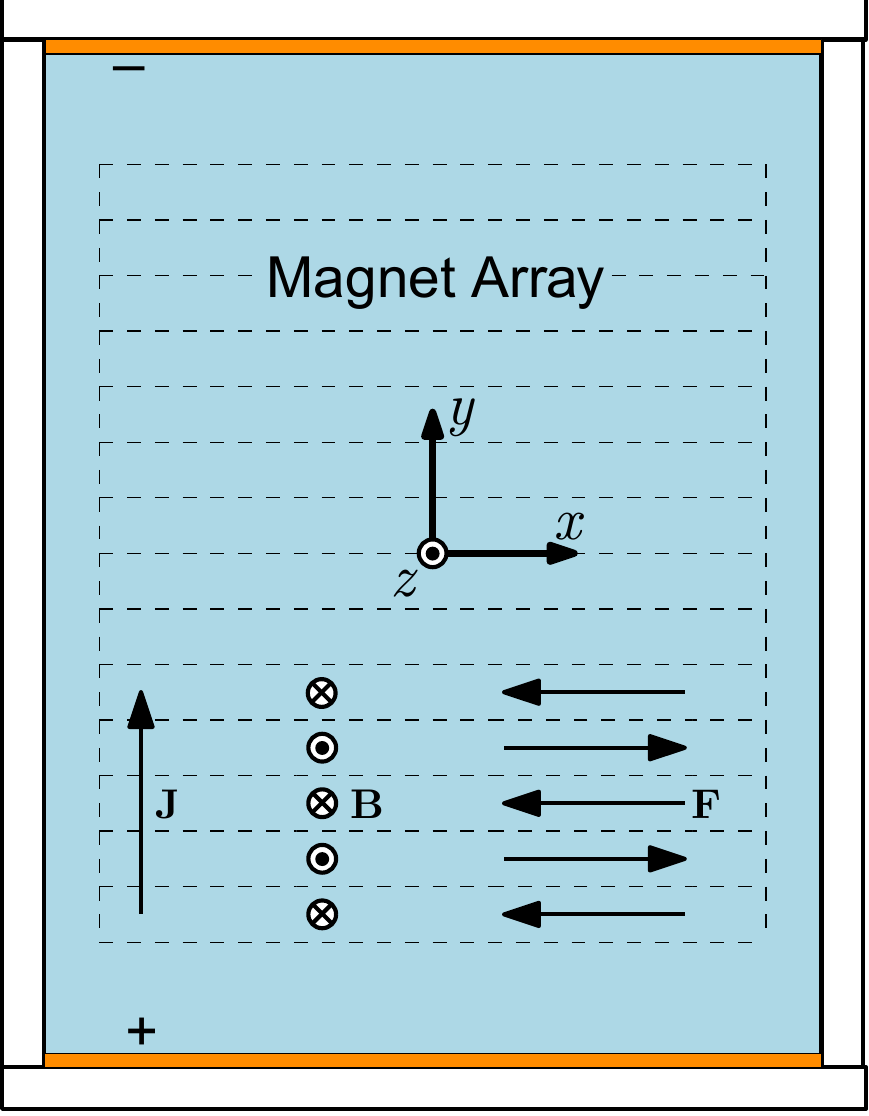}}\hspace{1mm}
\subfloat[]{\includegraphics[width=1.3in,valign=c]{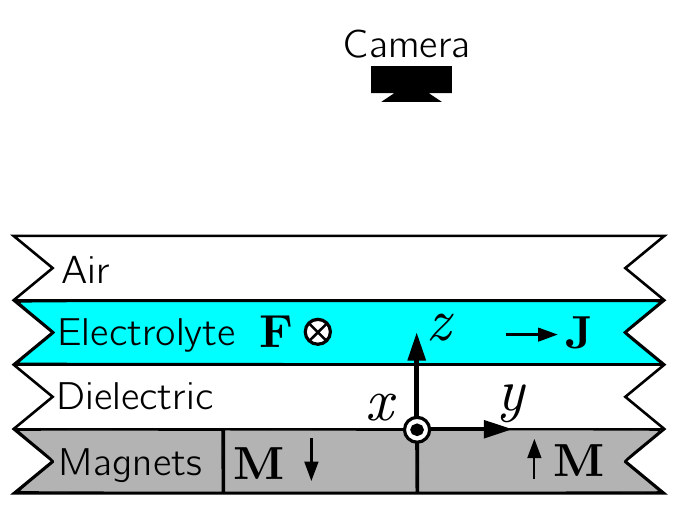}}
\caption{\label{fig:setup} {Experimental setup to generate Q2D Kolmogorov-like flow (a) Top view indicating  magnet array (dashed lines) and  directions of magnetic field ${\bf B}$, current density ${\bf J} =J\uv{y}$, and  electromagnetic forcing ${\bf F}$. (b) Side view showing  stably stratified {immiscible two-layer} configuration.}}
\end{figure}

A Q2D Kolmogorov-like flow in the experiment is generated in a shallow (6-mm thick) electrolyte-dielectric bilayer. The fluids lie in a {rectangular container with lateral ($x$ and $y$)} dimensions 17.8 cm $\times$ 22.9 cm (see \reffig{setup}). 
An array of permanent magnets placed beneath the container generates a near-sinusoidal magnetic field ${\bf B} \sim e^{-\pi z/w} \sin(\pi y/w) \uv{z}$, where $w = 1.27$ cm is the width of each magnet.
Passing a direct current ($J\uv{y}$) through the electrolyte layer generates a Lorentz force $\mathbf{F} = J\uv{y} \times {\bf B} \sim e^{-\pi z/w}\sin(\pi y/w)\uv{x}$ that drives a horizontal flow.
The {electrolyte-dielectric interface} is seeded with glass microspheres and  spatiotemporally resolved 2D velocity fields ${\bf u}(x,y,t)$  that quantify the horizontal flow are measured using particle image velocimetry  \cite{prana}. Details of the experiment and DNS are provided in the supplemental material (SM) \cite{Note1}.

The Q2D flow in experiment is theoretically modeled using the {nondimensional} 2D equation \cite{suri_2014},
\begin{equation}\label{eq:2dns}
{\partial_t \mathbf{u}} + \beta \mathbf{u} \cdot \nabla \mathbf{u} = - \nabla p + \frac{1}{Re}(\nabla^2 \mathbf{u}-\gamma \mathbf{u}) + {\bf f},
\end{equation}
which is derived by averaging 3D Navier-Stokes equation in the $z$ direction. Here,  $\mathbf{u}(x,y,t)$ is assumed to be incompressible ($\nabla \cdot {\bf u} = 0$) and corresponds to the velocity field at the free surface in experiment. $p$ is analogous to kinematic pressure. The spatial forcing profile ${\bf f}$ is obtained by depth-averaging and normalizing the Lorentz force ${\bf F}$.
Prefactor $\beta = 0.8$ to the nonlinear term and $-\gamma {\bf u}$ ($\gamma = 3.86$) {capture the  effects due to the solid boundary at the bottom of the fluid layers.}  Reynolds number $Re$ is related to the strength of electromagnetic forcing and is the parameter used to control the complexity of  flow (cf.  SM).

DNS of {the} flow governed by \refeqs{2dns} was performed using a {second-order (in space and time)} finite difference code previously employed in Refs. \cite{suri_2017a, tithof_2017, suri_2019}. The dimensions of the computational domain (14$w\times 18w$), no-slip velocity boundary conditions, and electromagnetic forcing in the DNS {correspond to} those in the experiment,  facilitating direct quantitative comparison between the two. The 2D forcing profile ${\bf f}$ in the DNS is antisymmetric under the inversion transformation $\rsym(x,y)\rightarrow(-x,-y)$, i.e., $\rsym{\bf f} = -{\bf f}$. Hence, \refeqs{2dns} is equivariant under $\rsym$. This two-fold symmetry  {($\rsym^2=\iden$)} is, however, weakly broken in experiment due to imperfections.

The Kolmogorov-like flow {becomes} weakly turbulent {above} $Re \approx 18$. Results {presented in this study correspond to $Re = 23.5\pm 1.5$ in experiment.} {In the DNS, turbulent time series were generated for $Re\in[22.6,25.1]$ in steps of $\Delta Re = 0.5$.} The flow is chaotic for these $Re$, which was validated in DNS by computing the Lyapunov exponents using continuous Gram-Schmidt orthogonalization {(cf. SM)}  \cite{egolf_2002,karimi_2012}. The {corresponding} Kaplan-Yorke dimension is $D_{KY} \approx 12$ and the Lyapunov time is $\tau_l\approx$ 50 seconds. We analyzed a  $36000\tau_l$-long turbulent time series in the DNS and experiment  to detect signatures of UPOs.

Time-periodic flows are solutions to \refeqs{2dns} {that satisfy the condition} ${\bf u}_{po}(t'+T) = {\bf u}_{po}(t')$. Here,  $t'$ parametrizes time along the orbit with period $T>0$.  
Due to equivariance under $\rsym$, \refeqs{2dns} can also possess  ``pre-periodic'' solutions {such that} ${\bf u}_{po}(t'+T) = \rsym{\bf u}_{po}(t')$ \cite{willis_2013}.
However, it is not known {\it a priori} {whether} UPOs of either type exist for our choice of parameters ($Re,\beta, \gamma$) and, whether turbulent flow transiently {approaches such solutions.} 

{To identify signatures of UPOs,} we performed recurrence analysis on the turbulent time series from DNS by computing  \cite{chandler_2013,willis_2013}:
\begin{equation}\label{eq:rec}
r(t,\tau) = D_c^{-1}\min_g \|g{\bf u}(t) - {\bf u}(t+\tau)\|,\quad g = \{\rsym,\iden\}.
\end{equation}
Here, $\tau>0$ and $\|\cdot\|$ represents the $L_2$ norm. The normalization constant $D_c=\max_{t,\tau}\|\mathbf{u}(t) - \mathbf{u}(t+\tau)\|$ is the empirically estimated diameter of the chaotic set which ensures $r(t,\tau)\leq 1$. Low recurrence values $r(t,\tau)\ll 1$ indicate that turbulent flow fields, or their symmetry-related copies, at instants $t$ and $t+\tau$ are similar.  Therefore, during the interval $[t,t+\tau]$, the turbulent trajectory in state space is possibly shadowing an unstable periodic or pre-periodic orbit with period $T\approx \tau$. Initializing a  Newton-Krylov solver \cite{viswanath_2007} with 50 initial conditions ${\bf u}(t)$  that corresponding to  deep minima in recurrence ($r \leq 0.2$), we identified seven {distinct} UPOs, labeled as follows:  \PkD{0}{}, \PkD{1}{}, \PkD{2}{A}, \PkD{2}{B}, \PkD{2}{C}, \PkD{3}{A} and \PkD{3}{B}. 
Among these,  \PkD{0}{} and \PkD{1}{}  are $\rsym-$invariant and have been reported previously \cite{suri_2019}. The rest lie in full state space; 
\PkD{2}{A-2C} are pre-periodic orbits {that lie on the same solution branch}
and \PkD{3}{B} is the symmetry-related copy of \PkD{3}{A}.  Several properties of the UPOs are tabulated in the SM.

\begin{figure}[!t]
	\centering
	\subfloat[]{\includegraphics[height=1.5in,valign=b]{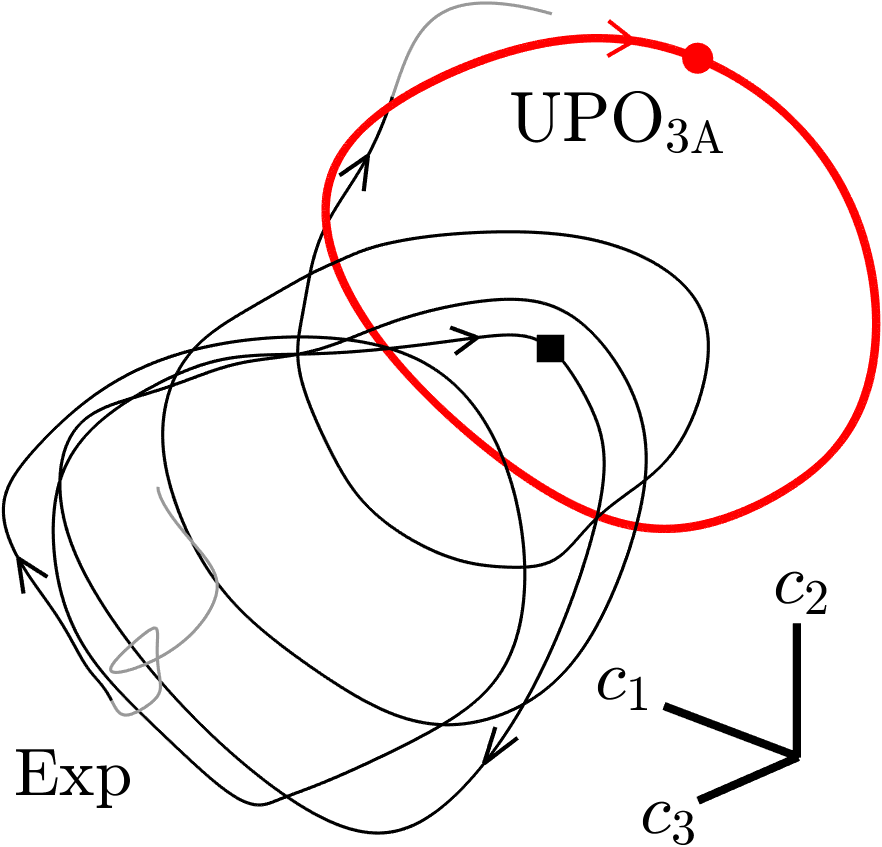}}\hspace{0mm}
	\subfloat[Exp]{\includegraphics[height=1in,valign=b]{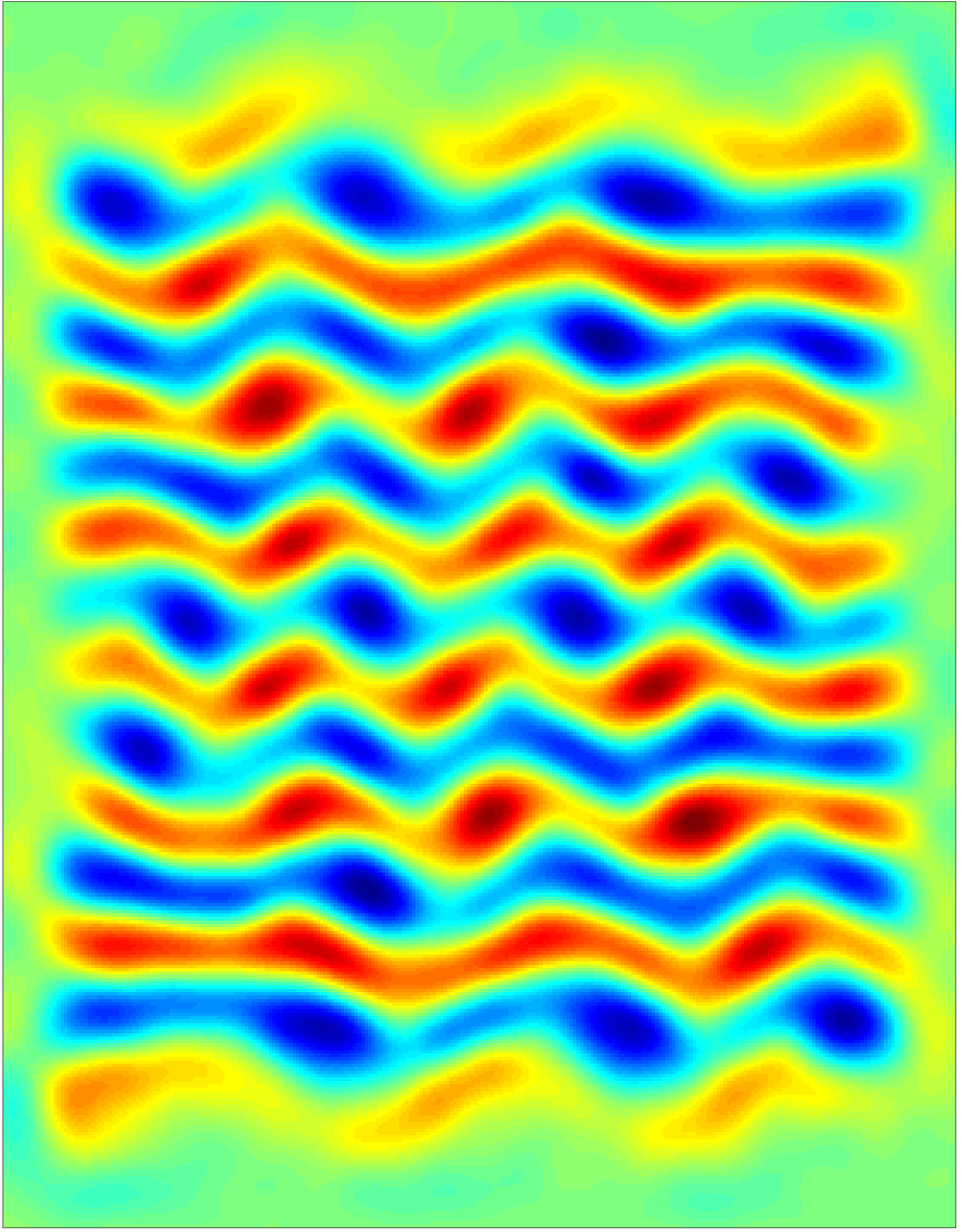}}\hspace{0mm}
	\subfloat[\PkD{3}{A}]{\includegraphics[height=1in,valign=b]{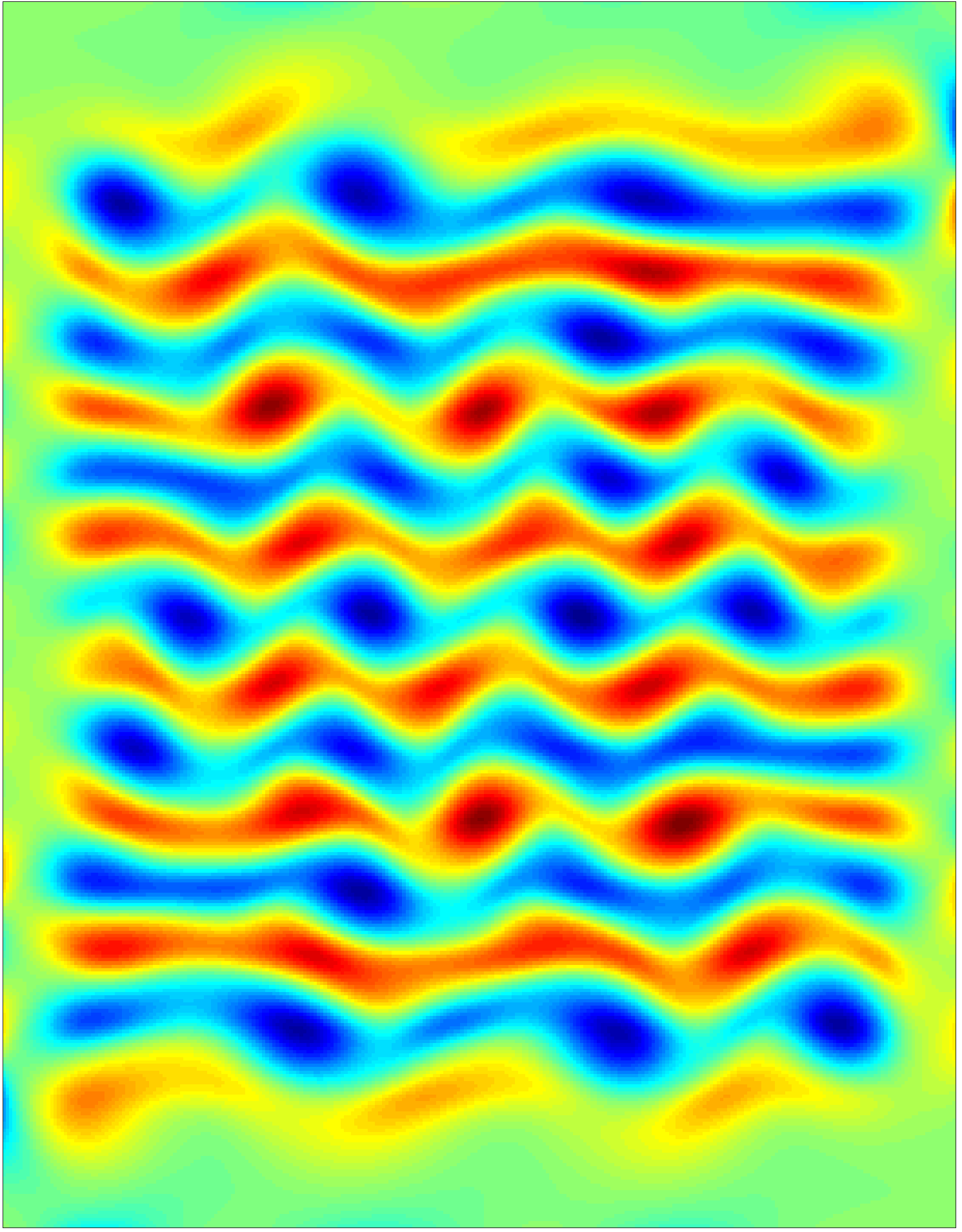}}
	\caption{\label{fig:compare} (a) Low-dimensional projection of state space showing turbulent trajectory from experiment (black curve) shadowing \PkD{3}{A} (red loop). {Each point on these curves represents a flow field}. The segment in black (gray) lies  in (outside) the neighborhood of \PkD{3}{A}.  The sphere and square indicate  instantaneously  closest points on the turbulent trajectory and \PkD{3}{A}. The corresponding flow snapshots are shown in (b) and (c), where color represents  vorticity $\omega = (\nabla \times {\bf u})_z$. The projection method is detailed in the SM.}	
\end{figure}

To test the dynamical relevance of a UPO in experiment, i.e., whether turbulent flows ${\bf u}(t)$ approach the UPO, we computed the normalized distance \cite{suri_2018, suri_2019}
\begin{equation}\label{eq:D1}
D_1(t) = D_c^{-1}\min\limits_{t'}\|{\bf u}(t) - {\bf u}_{po}(t')\|.
\end{equation}
$D_1$ is the instantaneous separation between ${\bf u}(t)$ and the closest point on the orbit ${\bf u}_{po}(t')$,  
{as shown in \reffig{compare}(a).}
$D_1 \ll 1$ ($D_1\approx 1$) implies the turbulent flow is very close to (far away from) the UPO in state space.
{We previously identified that flow fields in physical space are visually similar when $D_1\leq 0.45$ \cite{suri_2018}.} 
Using this metric, we found many instances when turbulent flow approaches one of the computed UPOs. 
For example, \reffig{compare} compares snapshots from experiment and \PkD{3}{A}  at an instant the turbulent trajectory is near \PkD{3}{A} ($D_1=0.16$). 
The remarkable similarity between these flow fields confirms that  turbulent trajectories in experiment indeed approach UPOs very closely. 

\begin{figure}[!t]
	\centering
	{\includegraphics[width=3.3in]{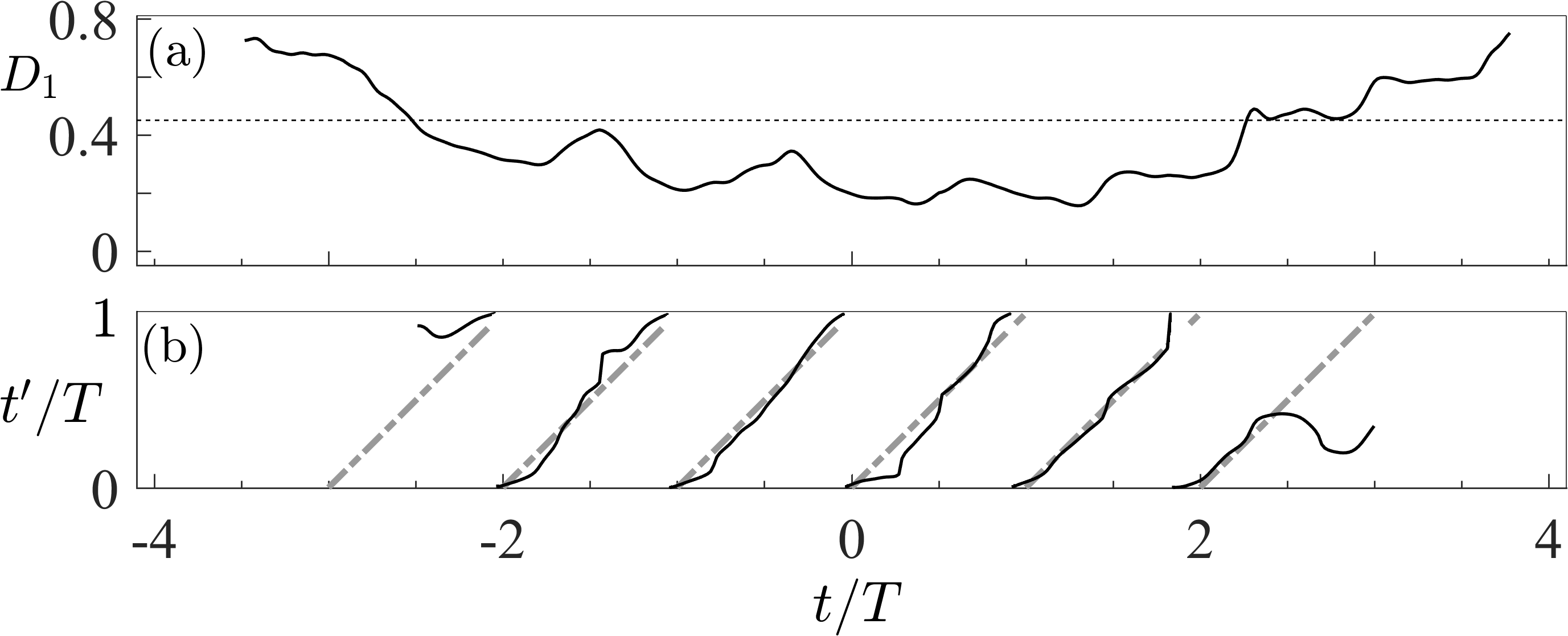}}\caption{\label{fig:D1} {(a) Instantaneous normalized separation $D_1$ between a turbulent trajectory in experiment and periodic orbit \PkD{3}{A}. {The dashed black line ($D_1 = 0.45$) indicates the limit for} closeness in state space. (b)  $t'$ and $t$ parametrize time along \PkD{3}{A} and the turbulent trajectory, respectively.}}
\end{figure}

Turbulent trajectories near a UPO should shadow its evolution in state space \cite{kawahara_2001,viswanath_2007,chandler_2013,budanur_2017a}.
To validate this in experiment, we analyzed a {particularly} close pass to \PkD{3}{A}; the  period  of this orbit is $T = 113.2\,$s ($2.2\tau_l$). 
Using {our} closeness criterion, we estimated that the turbulent trajectory {remains} in the neighborhood of 
\PkD{3}{A} for a duration equal to {about} four periods of \PkD{3}{A} ($-2<t/T<2$ in \reffig{D1}). 
To visualize turbulent dynamics {over} this interval,  we projected the state space around \PkD{3}{A} onto a low-dimensional subspace in \reffig{compare}. 
Indeed, the turbulent trajectory approaches \PkD{3}{A}, shadows its evolution by tracing four loops, and subsequently departs from the neighborhood of \PkD{3}{A}. Video 1 in the SM shows side-by-side comparison of  turbulent flow and \PkD{3}{A} in both physical space and state space.

Since the shapes of the turbulent trajectory and  \PkD{3}{A} are similar, one may ask if the corresponding flows evolve at similar rates.
To explore this, for each point on the turbulent trajectory ${\bf u}(t)$, we identified the closest point  ${\bf u}_{po}(t')$ on \PkD{3}{A} (cf. \reffig{compare}).   
We then {tested} whether the time $t'$  
{increases} at the same rate as $t$;  $dt'/dt = 1$ implies identical rates of evolution {for the turbulent flow and the UPO it is shadowing.}    
\reffig{D1}(b) shows {the relation between $t$ and $t'$ during the interval of shadowing.} 
{We defined $t'$ on the interval $0<t'<T$ due to periodicity of the \PkD{}{}.}
For each of the four periods, $t'$ (solid black line) follows the ``diagonal''  $t\,\mathrm{mod}\,T$ (dashed gray line). This shows the turbulent trajectory and \PkD{3}{A}  evolve at comparable rates, on average. 
Noticeable difference in the instantaneous rates of evolution is related to  turbulent trajectories not approaching \PkD{3}{A} infinitesimally {closely} \cite{suri_2018}.
We also found that turbulent trajectories in experiment  shadow \PkD{0}{} and \PkD{2}{B} 
for a duration that is nearly one and three times their respective periods (see Fig. S2 and S3 in the SM). 

\begin{figure}[!t]
	\centering
	\includegraphics[width=3in,valign=c]{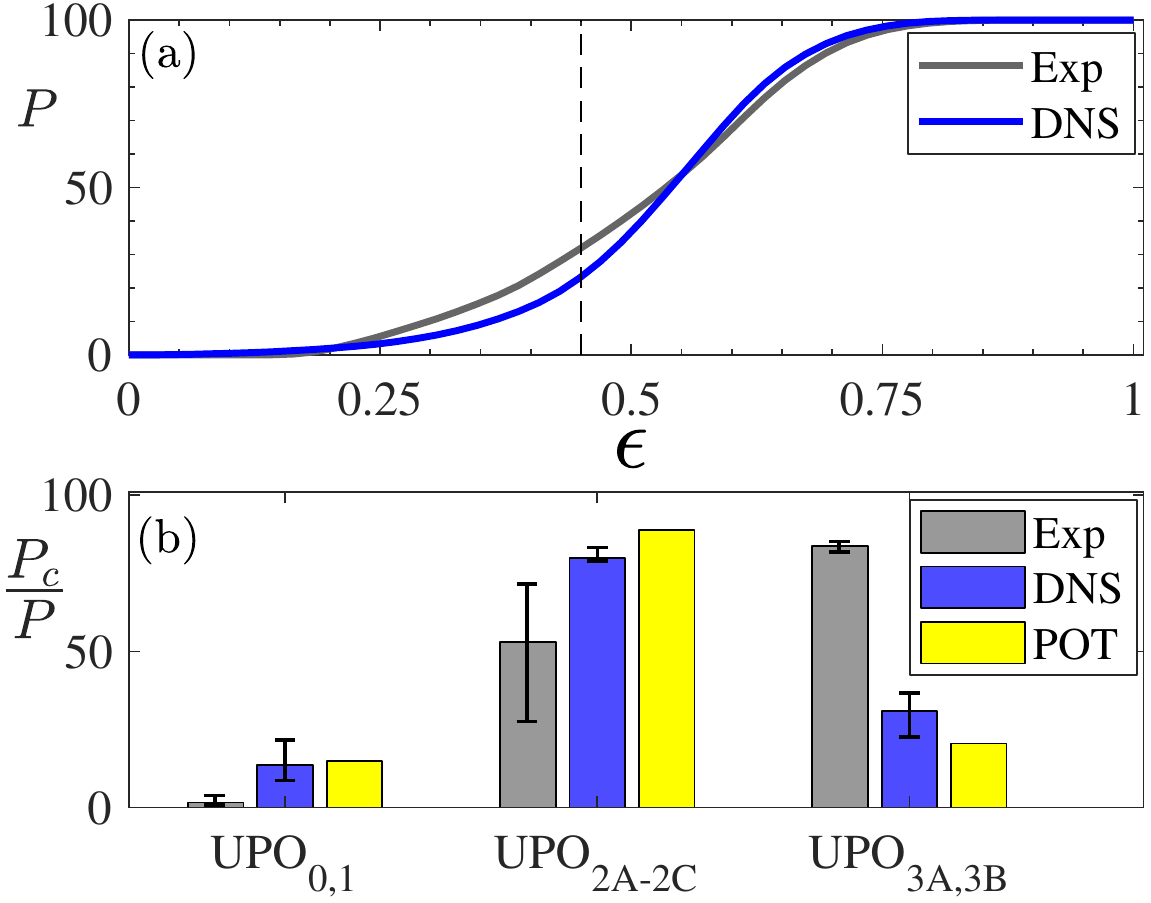}
	\caption{\label{fig:stats} {Statistical significance of UPOs. (a) Probability (in \%) 
	to find a turbulent trajectory at a normalized distance $D_1\leq\epsilon$ from the UPOs we computed. Dashed line indicates the upper limit $\epsilon = 0.45$ for closeness in state space. (b) Conditional  probabilities  for turbulent trajectories visiting  neighborhoods ($\epsilon = 0.45$) of UPO clusters. Error bars indicate changes to probabilities when  $\epsilon$ is varied between $[0.4,0.5]$. 
	}}
\end{figure}

{Statistical significance of UPOs has  received little attention in previous numerical studies \cite{kazantsev_1998,lucas_2015}, and   {none} in experiments. To address this, we computed the fraction $P(\epsilon)$ of the total time turbulent trajectories visit the $\epsilon-$neighborhood ($D_1\leq \epsilon$) of any UPO.  \reffig{stats}(a) reveals that  particularly close passes $(\epsilon \leq 0.2)$  to UPOs are rare ($P < 2\%$) and require very long turbulent time series for their detection. However, increasing the size of neighborhoods to $\epsilon = 0.45$,  we find that turbulent trajectories  spend a sizable fraction of time near  UPOs; about 30\% in experiment and 23\% in the DNS.
The sensitivity of $P$ to the choice of  $\epsilon$ is comparable to that observed by Kerswell \etal  for the statistical significance of traveling wave solutions in  turbulent pipe flow at $Re=2400$ \cite{kerswell_2007}.}

 {Since very close passes to UPOs are rare, quantifying the relative importance of various UPOs required coarse partitioning of the state space.}  
 A turbulent trajectory can be simultaneously close to several  UPOs which are adjacent to each other in state space. To  distinguish their statistical significance, we grouped the  UPOs into three clusters which are sufficiently far apart in state space:  \PkD{0}{,1}, \PkD{2}{A-2C}, and \PkD{3}{A,3B}. These clusters were identified using pairwise separation between UPOs (cf. SM). {For each cluster, we  then computed the  conditional probability  $P_c(\epsilon)/P(\epsilon)$  that a turbulent trajectory is near the UPOs in that cluster ($D_1\leq \epsilon$), given  it is near one of the seven UPOs.}
 
{The probabilities for turbulent trajectories in experiment and DNS visiting the three  UPO clusters are shown in \reffig{stats}(b) for $\epsilon = 0.45$. 
Clearly, the $\rsym$-invariant solutions \PkD{0}{,1} are rarely visited.
In contrast, UPO clusters that do not lie in the symmetry subspace are visited frequently and hence are statistically significant. Changing the neighborhood size between $\epsilon = 0.4$ and $\epsilon = 0.5$ did not affect the results qualitatively. 
The discrepancy between experiments and DNS appears to be a limitation of the 2D model in reproducing some aspects of an inherently 3D laboratory flow sufficiently accurately \cite{tithof_2017}. }

{The relative significance of UPO clusters can be rationalized using periodic orbit theory, originally developed for uniformly hyperbolic low-dimensional chaotic systems \cite{auerbach_1987, cvitanovic_1988}. 
The statistical ``weight'' associated with a UPO, and hence the probability of finding a chaotic trajectory  in its infinitesimal neighborhood, is  approximately given by (cf. Section 2.7.1 in Ref. \cite{lan_2010})
\begin{equation}\label{eq:weights}
\pi_i \propto \frac{1}{|\Lambda_{i1}|\cdot|\Lambda_{i2}|\cdots|\Lambda_{ik}|},
\end{equation}
where $|\Lambda_{i1}|$, $\cdots$, $|\Lambda_{ik}|$ are the magnitudes of the unstable Floquet multipliers of \PkD{i}{}. The POT weight associated with each cluster is then $P_c/P = \sum_i \pi_i$, where the summation is over the UPOs in that cluster. 
The weights $\pi_i$ in \refeqs{weights} are defined to within a normalization constant, which we chose such that the cumulative probability for the three clusters is the same for POT and DNS.
\reffig{stats}(b) shows that the statistical significance of various UPO clusters predicted  using POT is fairly consistent with measurements in DNS. 
This is quite remarkable, given that turbulent trajectories do not visit these UPOs infinitesimally closely.} {Lastly, alternative weighting formulas discussed in Refs. \cite{zoldi_1998,zoldi_1998b,kazantsev_1998,chandler_2013} also yield similar estimates for the statistical significance of UPO clusters (cf.  SM).}

\begin{figure}[!t]
	\includegraphics[width=3.1in]{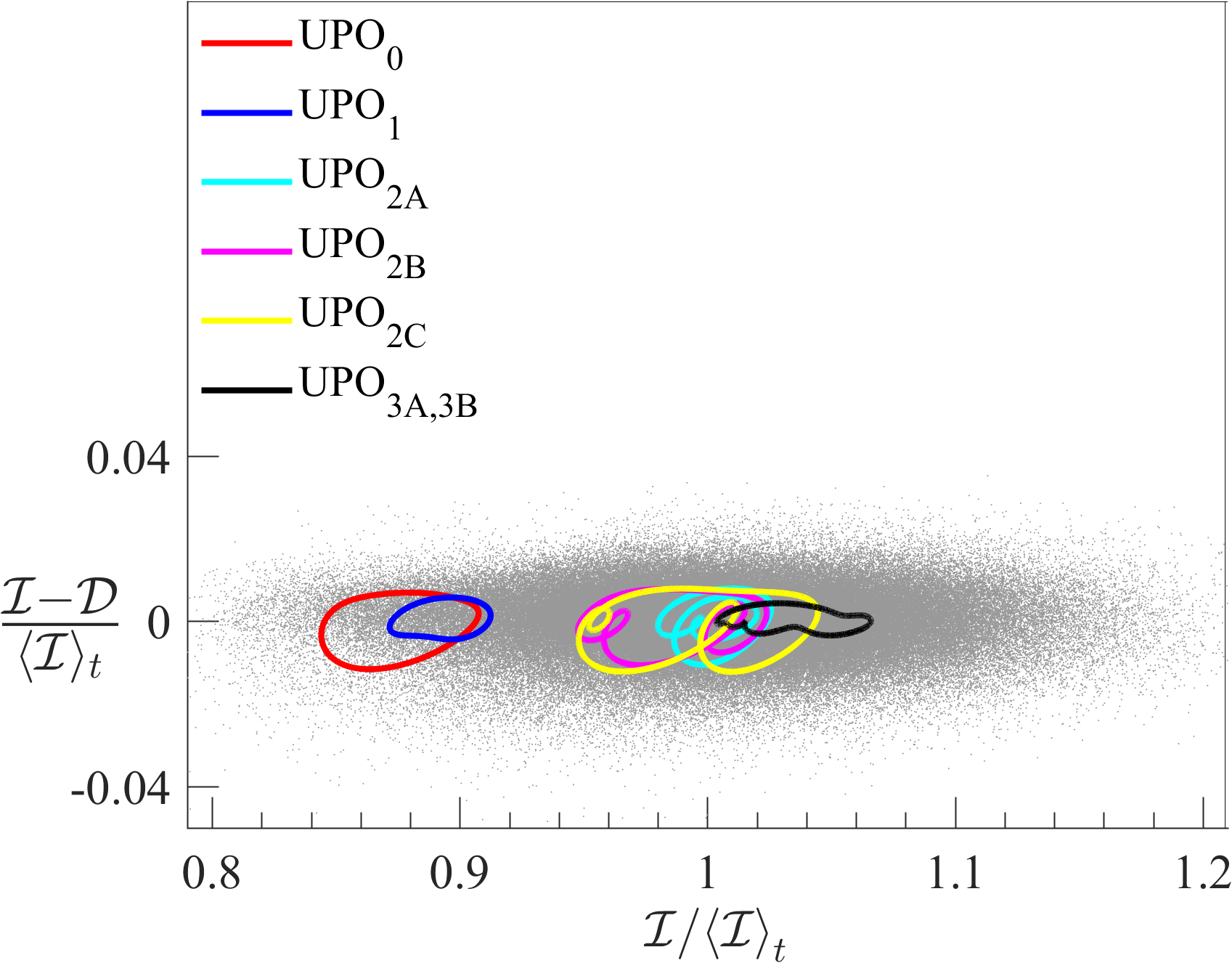}
	\begin{picture}(0,0)
	\put(-135,86){{\includegraphics[width=04.25cm]{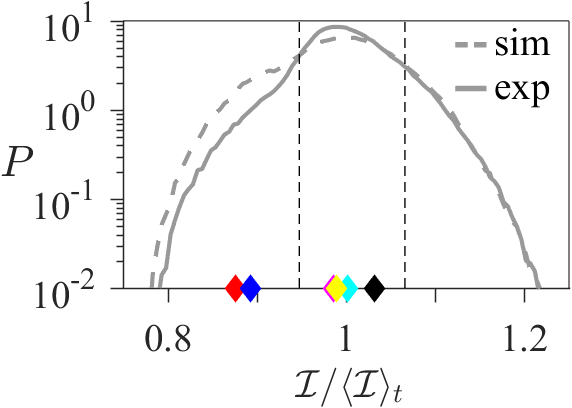}}}
	\end{picture}
	\caption{\label{fig:IvsD} 
		Energy input rate $\inp$ versus the difference  between input and dissipation  rates ($\inp-\dis$) for turbulent time series in experiment (scatter plot) and UPOs (closed loops). (Inset) Probability density function of $\inp(t)$ for turbulent flow in experiment (solid gray) and DNS (dashed gray). {Colored symbols show the mean values of $\inp$ for each of the seven UPOs and the dashed black lines represent the range of $\inp$ for \PkD{2}{A-2C} and \PkD{3}{A,3B}.}}
\end{figure}

{The motivation behind identifying UPOs and quantifying their statistical  significance is to compare  statistical averages of turbulent flows with those of UPOs.} 
Following standard practice \cite{kawahara_2001,viswanath_2007,vanVeen_2019}, we computed the instantaneous  energy input ($\inp$) and dissipation ($\dis$) rates
\begin{align}\label{eq:I}
\mathcal{I}(t) &= \tribra{{\bf f}\cdot{\bf u}}_{\Omega},\nonumber\\
\mathcal{D}(t) &= -\frac{1}{Re} \tribra{{\bf u}\cdot{\nabla^2 \bf u}-\gamma\, {\bf u}\cdot{\bf u}}_\Omega,
\end{align}
for the turbulent flow and all the UPOs. Here, ${\bf a}\cdot {\bf b}$ is the  scalar product between vector fields ${\bf a}$, ${\bf b}$ and $\tribra{\cdot}_\Omega$ represents  the integral $\int_\Omega(\cdots)dx\,dy$ evaluated over the entire flow domain $\Omega$. In \reffig{IvsD}, we plotted the difference between instantaneous input and dissipation rates ($\inp-\dis$) versus the energy input rate $\inp$ for the turbulent flow in experiment. $\inp$ and $\dis$ are normalized by the temporal mean $\tribra{\inp}_t = \tribra{\dis}_t$. 
The corresponding quantities for each UPO  are overlaid. Additionally, the probability density function for $\inp$ from experiment (as well as DNS) is shown in the inset. 

{For the statistically significant \PkD{2}{A-2C} and \PkD{3}{A,3B}}, both energy input and  dissipation rates cluster around the turbulent mean values,  located at $\inp/\tribra{\inp}_t = 1$ and ${\inp}-{\dis} = 0$ in \reffig{IvsD}.
 The $\inp$ (and $\dis$) values for these UPOs vary over a narrow range  (0.95,1.07) that is {approximately $\pm \sigma_{\inp}$ of the turbulent mean, where $\sigma_{\inp}=0.055$ is the standard deviation of $\inp$ for turbulent flow.} 
Consequently, the mean energy input (and dissipation) rate for each {of} these five UPOs is within {$\pm0.6\sigma_{\inp}$} of the turbulent average (unity), as shown in the inset.
{In contrast, \PkD{0}{,1}, which are statistically insignificant, have mean values of $\inp$ and $\dis$ that deviate by over $2\sigma_{\inp}$ from the turbulent mean value.}

In this article, we provided unambiguous experimental evidence for the {dynamical relevance and statistical significance} of UPOs in a moderately turbulent flow. 
{We  showed that turbulent trajectories in state space transiently approach UPOs closely and shadow their spatiotemporal evolution.}
We also quantified the statistical significance of various UPOs by computing the fraction of time turbulent trajectories visit their neighborhoods. {The estimates from  DNS are  consistent with the ``weights'' predicted by periodic orbit theory.}
Lastly, we showed that statistically significant UPOs capture  time-averaged properties of the turbulent flows in both experiment and DNS  accurately.

Our study identified that turbulent flows  spend about 30\% of the time near the UPOs we computed.  
This suggests that {UPOs with longer periods as well as other types of nonchaotic solutions--such as unstable equilibria, quasi-periodic orbits, and hetero/homoclinic connections--may also play an important dynamical and statistical role} \cite{schneider_2007a, vanveen_2011b, avila_2013,  farano_2018a}. 
Their  existence and dynamical relevance, at least in  symmetry-invariant subspaces,  was recently demonstrated for both 2D and three-dimensional shear flows \cite{schneider_2007a, avila_2013, farano_2018a, suri_2019}. 
{Hence, a dynamical framework based on  UPOs, as well as other types of recurrent solutions,  should ultimately enable forecasting \cite{suri_2017a, suri_2018} and control (e.g., L\"uthje \etal \cite{luthje_2001}) of turbulent dynamics,  besides accurately predicting its statistical properties.} 

\begin{acknowledgments}
	MS and RG acknowledge funding from the National Science Foundation (CMMI-1234436, DMS-1125302, CMMI-1725587) and Defense Advanced Research Projects Agency (HR0011-16-2-0033). B.S. acknowledges funding from the European Union’s Horizon 2020 research and innovation program under the Marie Sk\l{}odowska-Curie grant agreement No 754411.\\
\end{acknowledgments}


\begin{thebibliography}{43}%
	\makeatletter
	\providecommand \@ifxundefined [1]{%
		\@ifx{#1\undefined}
	}%
	\providecommand \@ifnum [1]{%
		\ifnum #1\expandafter \@firstoftwo
		\else \expandafter \@secondoftwo
		\fi
	}%
	\providecommand \@ifx [1]{%
		\ifx #1\expandafter \@firstoftwo
		\else \expandafter \@secondoftwo
		\fi
	}%
	\providecommand \natexlab [1]{#1}%
	\providecommand \enquote  [1]{``#1''}%
	\providecommand \bibnamefont  [1]{#1}%
	\providecommand \bibfnamefont [1]{#1}%
	\providecommand \citenamefont [1]{#1}%
	\providecommand \href@noop [0]{\@secondoftwo}%
	\providecommand \href [0]{\begingroup \@sanitize@url \@href}%
	\providecommand \@href[1]{\@@startlink{#1}\@@href}%
	\providecommand \@@href[1]{\endgroup#1\@@endlink}%
	\providecommand \@sanitize@url [0]{\catcode `\\12\catcode `\$12\catcode
		`\&12\catcode `\#12\catcode `\^12\catcode `\_12\catcode `\%12\relax}%
	\providecommand \@@startlink[1]{}%
	\providecommand \@@endlink[0]{}%
	\providecommand \url  [0]{\begingroup\@sanitize@url \@url }%
	\providecommand \@url [1]{\endgroup\@href {#1}{\urlprefix }}%
	\providecommand \urlprefix  [0]{URL }%
	\providecommand \Eprint [0]{\href }%
	\providecommand \doibase [0]{http://dx.doi.org/}%
	\providecommand \selectlanguage [0]{\@gobble}%
	\providecommand \bibinfo  [0]{\@secondoftwo}%
	\providecommand \bibfield  [0]{\@secondoftwo}%
	\providecommand \translation [1]{[#1]}%
	\providecommand \BibitemOpen [0]{}%
	\providecommand \bibitemStop [0]{}%
	\providecommand \bibitemNoStop [0]{.\EOS\space}%
	\providecommand \EOS [0]{\spacefactor3000\relax}%
	\providecommand \BibitemShut  [1]{\csname bibitem#1\endcsname}%
	\let\auto@bib@innerbib\@empty
	\bibitem [{\citenamefont {Hof}\ \emph {et~al.}(2004)\citenamefont {Hof},
		\citenamefont {van Doorne}, \citenamefont {Westerweel}, \citenamefont
		{Nieuwstadt}, \citenamefont {Faisst}, \citenamefont {Eckhardt}, \citenamefont
		{Wedin}, \citenamefont {Kerswell},\ and\ \citenamefont {Waleffe}}]{hof_2004}%
	\BibitemOpen
	\bibfield  {author} {\bibinfo {author} {\bibfnamefont {B.}~\bibnamefont
			{Hof}}, \bibinfo {author} {\bibfnamefont {C.~W.~H.}\ \bibnamefont {van
				Doorne}}, \bibinfo {author} {\bibfnamefont {J.}~\bibnamefont {Westerweel}},
		\bibinfo {author} {\bibfnamefont {F.~T.~M.}\ \bibnamefont {Nieuwstadt}},
		\bibinfo {author} {\bibfnamefont {H.}~\bibnamefont {Faisst}}, \bibinfo
		{author} {\bibfnamefont {B.}~\bibnamefont {Eckhardt}}, \bibinfo {author}
		{\bibfnamefont {H.}~\bibnamefont {Wedin}}, \bibinfo {author} {\bibfnamefont
			{R.~R.}\ \bibnamefont {Kerswell}}, \ and\ \bibinfo {author} {\bibfnamefont
			{F.}~\bibnamefont {Waleffe}},\ }\href@noop {} {\bibfield  {journal} {\bibinfo
			{journal} {Science}\ }\textbf {\bibinfo {volume} {305}},\ \bibinfo {pages}
		{1594} (\bibinfo {year} {2004})}\BibitemShut {NoStop}%
	\bibitem [{\citenamefont {Robinson}(1991)}]{robinson_1991}%
	\BibitemOpen
	\bibfield  {author} {\bibinfo {author} {\bibfnamefont {S.~K.}\ \bibnamefont
			{Robinson}},\ }\href@noop {} {\bibfield  {journal} {\bibinfo  {journal}
			{Annual Review of Fluid Mechanics}\ }\textbf {\bibinfo {volume} {23}},\
		\bibinfo {pages} {601} (\bibinfo {year} {1991})}\BibitemShut {NoStop}%
	\bibitem [{\citenamefont {Dennis}\ and\ \citenamefont
		{Sogaro}(2014)}]{dennis_2014}%
	\BibitemOpen
	\bibfield  {author} {\bibinfo {author} {\bibfnamefont {D.~J.~C.}\
			\bibnamefont {Dennis}}\ and\ \bibinfo {author} {\bibfnamefont {F.~M.}\
			\bibnamefont {Sogaro}},\ }\href@noop {} {\bibfield  {journal} {\bibinfo
			{journal} {Phys. Rev. Lett.}\ }\textbf {\bibinfo {volume} {113}},\ \bibinfo
		{pages} {234501} (\bibinfo {year} {2014})}\BibitemShut {NoStop}%
	\bibitem [{\citenamefont {Boffetta}\ and\ \citenamefont
		{Ecke}(2012)}]{boffetta_2012}%
	\BibitemOpen
	\bibfield  {author} {\bibinfo {author} {\bibfnamefont {G.}~\bibnamefont
			{Boffetta}}\ and\ \bibinfo {author} {\bibfnamefont {R.~E.}\ \bibnamefont
			{Ecke}},\ }\href@noop {} {\bibfield  {journal} {\bibinfo  {journal} {Annu.
				Rev. Fluid Mech.}\ }\textbf {\bibinfo {volume} {44}},\ \bibinfo {pages} {427}
		(\bibinfo {year} {2012})}\BibitemShut {NoStop}%
	\bibitem [{\citenamefont {{van Veen}}\ \emph {et~al.}(2019)\citenamefont {{van
				Veen}}, \citenamefont {{Vela-Martin}},\ and\ \citenamefont
		{{Kawahara}}}]{vanVeen_2019}%
	\BibitemOpen
	\bibfield  {author} {\bibinfo {author} {\bibfnamefont {L.}~\bibnamefont {{van
					Veen}}}, \bibinfo {author} {\bibfnamefont {A.}~\bibnamefont {{Vela-Martin}}},
		\ and\ \bibinfo {author} {\bibfnamefont {G.}~\bibnamefont {{Kawahara}}},\
	}\href@noop {} {\bibfield  {journal} {\bibinfo  {journal} {Phys. Rev. Lett.}\
		} (\bibinfo {year} {2019})}\BibitemShut {NoStop}%
	\bibitem [{\citenamefont {Goto}(2012)}]{goto_2012}%
	\BibitemOpen
	\bibfield  {author} {\bibinfo {author} {\bibfnamefont {S.}~\bibnamefont
			{Goto}},\ }\href@noop {} {\bibfield  {journal} {\bibinfo  {journal} {Progress
				of Theoretical Physics Supplement}\ }\textbf {\bibinfo {volume} {195}},\
		\bibinfo {pages} {139} (\bibinfo {year} {2012})}\BibitemShut {NoStop}%
	\bibitem [{\citenamefont {Kawahara}\ and\ \citenamefont
		{Kida}(2001)}]{kawahara_2001}%
	\BibitemOpen
	\bibfield  {author} {\bibinfo {author} {\bibfnamefont {G.}~\bibnamefont
			{Kawahara}}\ and\ \bibinfo {author} {\bibfnamefont {S.}~\bibnamefont
			{Kida}},\ }\href@noop {} {\bibfield  {journal} {\bibinfo  {journal} {J. Fluid
				Mech.}\ }\textbf {\bibinfo {volume} {449}},\ \bibinfo {pages} {291} (\bibinfo
		{year} {2001})}\BibitemShut {NoStop}%
	\bibitem [{\citenamefont {Gibson}\ \emph {et~al.}(2008)\citenamefont {Gibson},
		\citenamefont {Halcrow},\ and\ \citenamefont {Cvitanovi{\'c}}}]{gibson_2008}%
	\BibitemOpen
	\bibfield  {author} {\bibinfo {author} {\bibfnamefont {J.~F.}\ \bibnamefont
			{Gibson}}, \bibinfo {author} {\bibfnamefont {J.}~\bibnamefont {Halcrow}}, \
		and\ \bibinfo {author} {\bibfnamefont {P.}~\bibnamefont {Cvitanovi{\'c}}},\
	}\href@noop {} {\bibfield  {journal} {\bibinfo  {journal} {J. Fluid Mech.}\
		}\textbf {\bibinfo {volume} {611}},\ \bibinfo {pages} {107} (\bibinfo {year}
		{2008})}\BibitemShut {NoStop}%
	\bibitem [{\citenamefont {de~Lozar}\ \emph {et~al.}(2012)\citenamefont
		{de~Lozar}, \citenamefont {Mellibovsky}, \citenamefont {Avila},\ and\
		\citenamefont {Hof}}]{lozar_2012}%
	\BibitemOpen
	\bibfield  {author} {\bibinfo {author} {\bibfnamefont {A.}~\bibnamefont
			{de~Lozar}}, \bibinfo {author} {\bibfnamefont {F.}~\bibnamefont
			{Mellibovsky}}, \bibinfo {author} {\bibfnamefont {M.}~\bibnamefont {Avila}},
		\ and\ \bibinfo {author} {\bibfnamefont {B.}~\bibnamefont {Hof}},\
	}\href@noop {} {\bibfield  {journal} {\bibinfo  {journal} {Phys. Rev. Lett.}\
		}\textbf {\bibinfo {volume} {108}},\ \bibinfo {pages} {214502} (\bibinfo
		{year} {2012})}\BibitemShut {NoStop}%
	\bibitem [{\citenamefont {Avila}\ \emph {et~al.}(2013)\citenamefont {Avila},
		\citenamefont {Mellibovsky}, \citenamefont {Roland},\ and\ \citenamefont
		{Hof}}]{avila_2013}%
	\BibitemOpen
	\bibfield  {author} {\bibinfo {author} {\bibfnamefont {M.}~\bibnamefont
			{Avila}}, \bibinfo {author} {\bibfnamefont {F.}~\bibnamefont {Mellibovsky}},
		\bibinfo {author} {\bibfnamefont {N.}~\bibnamefont {Roland}}, \ and\ \bibinfo
		{author} {\bibfnamefont {B.}~\bibnamefont {Hof}},\ }\href {\doibase
		10.1103/PhysRevLett.110.224502} {\bibfield  {journal} {\bibinfo  {journal}
			{Phys. Rev. Lett.}\ }\textbf {\bibinfo {volume} {110}},\ \bibinfo {pages}
		{224502} (\bibinfo {year} {2013})}\BibitemShut {NoStop}%
	\bibitem [{\citenamefont {Suri}\ \emph {et~al.}(2017)\citenamefont {Suri},
		\citenamefont {Tithof}, \citenamefont {Grigoriev},\ and\ \citenamefont
		{Schatz}}]{suri_2017a}%
	\BibitemOpen
	\bibfield  {author} {\bibinfo {author} {\bibfnamefont {B.}~\bibnamefont
			{Suri}}, \bibinfo {author} {\bibfnamefont {J.}~\bibnamefont {Tithof}},
		\bibinfo {author} {\bibfnamefont {R.~O.}\ \bibnamefont {Grigoriev}}, \ and\
		\bibinfo {author} {\bibfnamefont {M.~F.}\ \bibnamefont {Schatz}},\ }\href
	{\doibase 10.1103/PhysRevLett.118.114501} {\bibfield  {journal} {\bibinfo
			{journal} {Phys. Rev. Lett.}\ }\textbf {\bibinfo {volume} {118}},\ \bibinfo
		{pages} {114501} (\bibinfo {year} {2017})}\BibitemShut {NoStop}%
	\bibitem [{\citenamefont {Hopf}(1948)}]{hopf_1948}%
	\BibitemOpen
	\bibfield  {author} {\bibinfo {author} {\bibfnamefont {E.}~\bibnamefont
			{Hopf}},\ }\href@noop {} {\bibfield  {journal} {\bibinfo  {journal} {Commun.
				Pur. Appl. Math.}\ }\textbf {\bibinfo {volume} {1}},\ \bibinfo {pages} {303}
		(\bibinfo {year} {1948})}\BibitemShut {NoStop}%
	\bibitem [{\citenamefont {Auerbach}\ \emph {et~al.}(1987)\citenamefont
		{Auerbach}, \citenamefont {Cvitanovi{\'c}}, \citenamefont {Eckmann},
		\citenamefont {Gunaratne},\ and\ \citenamefont {Procaccia}}]{auerbach_1987}%
	\BibitemOpen
	\bibfield  {author} {\bibinfo {author} {\bibfnamefont {D.}~\bibnamefont
			{Auerbach}}, \bibinfo {author} {\bibfnamefont {P.}~\bibnamefont
			{Cvitanovi{\'c}}}, \bibinfo {author} {\bibfnamefont {J.~P.}\ \bibnamefont
			{Eckmann}}, \bibinfo {author} {\bibfnamefont {G.}~\bibnamefont {Gunaratne}},
		\ and\ \bibinfo {author} {\bibfnamefont {I.}~\bibnamefont {Procaccia}},\
	}\href@noop {} {\bibfield  {journal} {\bibinfo  {journal} {Phys. Rev. Lett.}\
		}\textbf {\bibinfo {volume} {58}},\ \bibinfo {pages} {23} (\bibinfo {year}
		{1987})}\BibitemShut {NoStop}%
	\bibitem [{\citenamefont {Cvitanovi\ifmmode~\acute{c}\else
			\'{c}\fi{}}(1988)}]{cvitanovic_1988}%
	\BibitemOpen
	\bibfield  {author} {\bibinfo {author} {\bibfnamefont {P.}~\bibnamefont
			{Cvitanovi\ifmmode~\acute{c}\else \'{c}\fi{}}},\ }\href {\doibase
		10.1103/PhysRevLett.61.2729} {\bibfield  {journal} {\bibinfo  {journal}
			{Phys. Rev. Lett.}\ }\textbf {\bibinfo {volume} {61}},\ \bibinfo {pages}
		{2729} (\bibinfo {year} {1988})}\BibitemShut {NoStop}%
	\bibitem [{\citenamefont {Lan}(2010)}]{lan_2010}%
	\BibitemOpen
	\bibfield  {author} {\bibinfo {author} {\bibfnamefont {Y.}~\bibnamefont
			{Lan}},\ }\href@noop {} {\bibfield  {journal} {\bibinfo  {journal} {Commun.
				Nonlinear Sci.}\ }\textbf {\bibinfo {volume} {15}},\ \bibinfo {pages} {502}
		(\bibinfo {year} {2010})}\BibitemShut {NoStop}%
	\bibitem [{\citenamefont {Gibson}(2014)}]{channelflow}%
	\BibitemOpen
	\bibfield  {author} {\bibinfo {author} {\bibfnamefont {J.~F.}\ \bibnamefont
			{Gibson}},\ }\href@noop {} {\emph {\bibinfo {title} {{Channelflow}: {A}
				spectral {Navier-Stokes} simulator in {C}++}}},\ \bibinfo {type} {Tech.
		Rep.}\ (\bibinfo  {institution} {U. New Hampshire},\ \bibinfo {year} {2014})\
	\bibinfo {note} {{\tt {Channelflow.org}}}\BibitemShut {NoStop}%
	\bibitem [{\citenamefont {Chandler}\ and\ \citenamefont
		{Kerswell}(2013)}]{chandler_2013}%
	\BibitemOpen
	\bibfield  {author} {\bibinfo {author} {\bibfnamefont {G.~J.}\ \bibnamefont
			{Chandler}}\ and\ \bibinfo {author} {\bibfnamefont {R.~R.}\ \bibnamefont
			{Kerswell}},\ }\href@noop {} {\bibfield  {journal} {\bibinfo  {journal} {J.
				Fluid Mech.}\ }\textbf {\bibinfo {volume} {722}},\ \bibinfo {pages} {554}
		(\bibinfo {year} {2013})}\BibitemShut {NoStop}%
	\bibitem [{\citenamefont {Budanur}\ \emph {et~al.}(2017)\citenamefont
		{Budanur}, \citenamefont {Short}, \citenamefont {Farazmand}, \citenamefont
		{Willis},\ and\ \citenamefont {Cvitanović}}]{budanur_2017a}%
	\BibitemOpen
	\bibfield  {author} {\bibinfo {author} {\bibfnamefont {N.~B.}\ \bibnamefont
			{Budanur}}, \bibinfo {author} {\bibfnamefont {K.~Y.}\ \bibnamefont {Short}},
		\bibinfo {author} {\bibfnamefont {M.}~\bibnamefont {Farazmand}}, \bibinfo
		{author} {\bibfnamefont {A.~P.}\ \bibnamefont {Willis}}, \ and\ \bibinfo
		{author} {\bibfnamefont {P.}~\bibnamefont {Cvitanović}},\ }\href {\doibase
		10.1017/jfm.2017.699} {\bibfield  {journal} {\bibinfo  {journal} {J. Fluid
				Mech.}\ }\textbf {\bibinfo {volume} {833}},\ \bibinfo {pages} {274–301}
		(\bibinfo {year} {2017})}\BibitemShut {NoStop}%
	\bibitem [{\citenamefont {Suri}\ \emph {et~al.}(2019)\citenamefont {Suri},
		\citenamefont {Pallantla}, \citenamefont {Schatz},\ and\ \citenamefont
		{Grigoriev}}]{suri_2019}%
	\BibitemOpen
	\bibfield  {author} {\bibinfo {author} {\bibfnamefont {B.}~\bibnamefont
			{Suri}}, \bibinfo {author} {\bibfnamefont {R.~K.}\ \bibnamefont {Pallantla}},
		\bibinfo {author} {\bibfnamefont {M.~F.}\ \bibnamefont {Schatz}}, \ and\
		\bibinfo {author} {\bibfnamefont {R.~O.}\ \bibnamefont {Grigoriev}},\ }\href
	{\doibase 10.1103/PhysRevE.100.013112} {\bibfield  {journal} {\bibinfo
			{journal} {Phys. Rev. E}\ }\textbf {\bibinfo {volume} {100}},\ \bibinfo
		{pages} {013112} (\bibinfo {year} {2019})}\BibitemShut {NoStop}%
	\bibitem [{\citenamefont {Kostelich}\ \emph {et~al.}(1997)\citenamefont
		{Kostelich}, \citenamefont {Kan}, \citenamefont {Grebogi}, \citenamefont
		{Ott},\ and\ \citenamefont {Yorke}}]{kostelich1997}%
	\BibitemOpen
	\bibfield  {author} {\bibinfo {author} {\bibfnamefont {E.~J.}\ \bibnamefont
			{Kostelich}}, \bibinfo {author} {\bibfnamefont {I.}~\bibnamefont {Kan}},
		\bibinfo {author} {\bibfnamefont {C.}~\bibnamefont {Grebogi}}, \bibinfo
		{author} {\bibfnamefont {E.}~\bibnamefont {Ott}}, \ and\ \bibinfo {author}
		{\bibfnamefont {J.~A.}\ \bibnamefont {Yorke}},\ }\href@noop {} {\bibfield
		{journal} {\bibinfo  {journal} {Physica D: Nonlinear Phenomena}\ }\textbf
		{\bibinfo {volume} {109}},\ \bibinfo {pages} {81} (\bibinfo {year}
		{1997})}\BibitemShut {NoStop}%
	\bibitem [{\citenamefont {Toh}\ and\ \citenamefont {Itano}(2003)}]{toh_2003}%
	\BibitemOpen
	\bibfield  {author} {\bibinfo {author} {\bibfnamefont {S.}~\bibnamefont
			{Toh}}\ and\ \bibinfo {author} {\bibfnamefont {T.}~\bibnamefont {Itano}},\
	}\href {\doibase 10.1017/S0022112003003768} {\bibfield  {journal} {\bibinfo
			{journal} {J. Fluid Mech.}\ }\textbf {\bibinfo {volume} {481}},\ \bibinfo
		{pages} {67–76} (\bibinfo {year} {2003})}\BibitemShut {NoStop}%
	\bibitem [{\citenamefont {Viswanath}(2007)}]{viswanath_2007}%
	\BibitemOpen
	\bibfield  {author} {\bibinfo {author} {\bibfnamefont {D.}~\bibnamefont
			{Viswanath}},\ }\href@noop {} {\bibfield  {journal} {\bibinfo  {journal} {J.
				Fluid Mech.}\ }\textbf {\bibinfo {volume} {580}},\ \bibinfo {pages} {339}
		(\bibinfo {year} {2007})}\BibitemShut {NoStop}%
	\bibitem [{\citenamefont {Duguet}\ \emph {et~al.}(2008)\citenamefont {Duguet},
		\citenamefont {Pringle},\ and\ \citenamefont {Kerswell}}]{duguet_2008a}%
	\BibitemOpen
	\bibfield  {author} {\bibinfo {author} {\bibfnamefont {Y.}~\bibnamefont
			{Duguet}}, \bibinfo {author} {\bibfnamefont {C.~C.~T.}\ \bibnamefont
			{Pringle}}, \ and\ \bibinfo {author} {\bibfnamefont {R.~R.}\ \bibnamefont
			{Kerswell}},\ }\href@noop {} {\bibfield  {journal} {\bibinfo  {journal}
			{Phys. Fluids}\ }\textbf {\bibinfo {volume} {20}},\ \bibinfo {pages} {114102}
		(\bibinfo {year} {2008})}\BibitemShut {NoStop}%
	\bibitem [{\citenamefont {van Veen}\ and\ \citenamefont
		{Kawahara}(2011)}]{vanveen_2011b}%
	\BibitemOpen
	\bibfield  {author} {\bibinfo {author} {\bibfnamefont {L.}~\bibnamefont {van
				Veen}}\ and\ \bibinfo {author} {\bibfnamefont {G.}~\bibnamefont {Kawahara}},\
	}\href {\doibase 10.1103/PhysRevLett.107.114501} {\bibfield  {journal}
		{\bibinfo  {journal} {Phys. Rev. Lett.}\ }\textbf {\bibinfo {volume} {107}},\
		\bibinfo {pages} {114501} (\bibinfo {year} {2011})}\BibitemShut {NoStop}%
	\bibitem [{\citenamefont {Kreilos}\ and\ \citenamefont
		{Eckhardt}(2012)}]{kreilos_2012}%
	\BibitemOpen
	\bibfield  {author} {\bibinfo {author} {\bibfnamefont {T.}~\bibnamefont
			{Kreilos}}\ and\ \bibinfo {author} {\bibfnamefont {B.}~\bibnamefont
			{Eckhardt}},\ }\href@noop {} {\bibfield  {journal} {\bibinfo  {journal}
			{Chaos: An Interdisciplinary Journal of Nonlinear Science}\ }\textbf
		{\bibinfo {volume} {22}},\ \bibinfo {pages} {047505} (\bibinfo {year}
		{2012})}\BibitemShut {NoStop}%
	\bibitem [{\citenamefont {Willis}\ \emph {et~al.}(2013)\citenamefont {Willis},
		\citenamefont {Cvitanovi{\'c}},\ and\ \citenamefont {Avila}}]{willis_2013}%
	\BibitemOpen
	\bibfield  {author} {\bibinfo {author} {\bibfnamefont {A.~P.}\ \bibnamefont
			{Willis}}, \bibinfo {author} {\bibfnamefont {P.}~\bibnamefont
			{Cvitanovi{\'c}}}, \ and\ \bibinfo {author} {\bibfnamefont {M.}~\bibnamefont
			{Avila}},\ }\href@noop {} {\bibfield  {journal} {\bibinfo  {journal} {J.
				Fluid Mech.}\ }\textbf {\bibinfo {volume} {721}},\ \bibinfo {pages} {514}
		(\bibinfo {year} {2013})}\BibitemShut {NoStop}%
	\bibitem [{\citenamefont {Page}\ and\ \citenamefont
		{Kerswell}(2020)}]{page_2020}%
	\BibitemOpen
	\bibfield  {author} {\bibinfo {author} {\bibfnamefont {J.}~\bibnamefont
			{Page}}\ and\ \bibinfo {author} {\bibfnamefont {R.~R.}\ \bibnamefont
			{Kerswell}},\ }\href {\doibase 10.1017/jfm.2019.1074} {\bibfield  {journal}
		{\bibinfo  {journal} {Journal of Fluid Mechanics}\ }\textbf {\bibinfo
			{volume} {886}},\ \bibinfo {pages} {A28} (\bibinfo {year}
		{2020})}\BibitemShut {NoStop}%
	\bibitem [{\citenamefont {Waleffe}(1997)}]{waleffe_1997}%
	\BibitemOpen
	\bibfield  {author} {\bibinfo {author} {\bibfnamefont {F.}~\bibnamefont
			{Waleffe}},\ }\href@noop {} {\bibfield  {journal} {\bibinfo  {journal} {Phys.
				Fluids}\ }\textbf {\bibinfo {volume} {9}},\ \bibinfo {pages} {883} (\bibinfo
		{year} {1997})}\BibitemShut {NoStop}%
	\bibitem [{\citenamefont {Lucas}\ and\ \citenamefont
		{Kerswell}(2015)}]{lucas_2015}%
	\BibitemOpen
	\bibfield  {author} {\bibinfo {author} {\bibfnamefont {D.}~\bibnamefont
			{Lucas}}\ and\ \bibinfo {author} {\bibfnamefont {R.~R.}\ \bibnamefont
			{Kerswell}},\ }\href@noop {} {\bibfield  {journal} {\bibinfo  {journal}
			{Phys. Fluids}\ }\textbf {\bibinfo {volume} {27}},\ \bibinfo {pages} {045106}
		(\bibinfo {year} {2015})}\BibitemShut {NoStop}%
	\bibitem [{\citenamefont {Drew}\ \emph {et~al.}(2013)\citenamefont {Drew},
		\citenamefont {Charonko},\ and\ \citenamefont {Vlachos}}]{prana}%
	\BibitemOpen
	\bibfield  {author} {\bibinfo {author} {\bibfnamefont {B.}~\bibnamefont
			{Drew}}, \bibinfo {author} {\bibfnamefont {J.}~\bibnamefont {Charonko}}, \
		and\ \bibinfo {author} {\bibfnamefont {P.~P.}\ \bibnamefont {Vlachos}},\
	}\href@noop {} {\enquote {\bibinfo {title} {{QI} -- {Q}uantitative {I}maging
				({PIV} and more)},}\ } (\bibinfo {year} {2013}),\ \bibinfo {note} {available
		at https://sourceforge.net/projects/qi-tools/}\BibitemShut {NoStop}%
	\bibitem [{Note1()}]{Note1}%
	\BibitemOpen
	\bibinfo {note} {See supplemental material for details regarding (i)
		experimental setup, (ii) DNS, (iii) $D_{KY}$ computation, (iv) properties of
		UPOs, (v) state space projection procedure, (vi) turbulent trajectories
		shadowing UPO$_{{0}{\protect \textrm {}}}$ and UPO$_{{2}{\protect \textrm
				{B}}}$, (vii) pairwise separation between UPO$_{{}{\protect \textrm {}}}$s,
		and (viii) comparison of UPO weighting protocols. Videos 1, 2, and 3 show
		side-by-side comparison of turbulent flows in experiment shadowing
		UPO$_{{3}{\protect \textrm {A}}}$, UPO$_{{0}{\protect \textrm {}}}$, and
		UPO$_{{2}{\protect \textrm {B}}}$, respectively.}\BibitemShut {Stop}%
	\bibitem [{\citenamefont {Suri}\ \emph {et~al.}(2014)\citenamefont {Suri},
		\citenamefont {Tithof}, \citenamefont {Mitchell}, \citenamefont {Grigoriev},\
		and\ \citenamefont {Schatz}}]{suri_2014}%
	\BibitemOpen
	\bibfield  {author} {\bibinfo {author} {\bibfnamefont {B.}~\bibnamefont
			{Suri}}, \bibinfo {author} {\bibfnamefont {J.}~\bibnamefont {Tithof}},
		\bibinfo {author} {\bibfnamefont {R.}~\bibnamefont {Mitchell}}, \bibinfo
		{author} {\bibfnamefont {R.~O.}\ \bibnamefont {Grigoriev}}, \ and\ \bibinfo
		{author} {\bibfnamefont {M.~F.}\ \bibnamefont {Schatz}},\ }\href {\doibase
		http://dx.doi.org/10.1063/1.4873417} {\bibfield  {journal} {\bibinfo
			{journal} {Phys. Fluids}\ }\textbf {\bibinfo {volume} {26}},\ \bibinfo {eid}
		{053601} (\bibinfo {year} {2014})}\BibitemShut {NoStop}%
	\bibitem [{\citenamefont {Tithof}\ \emph {et~al.}(2017)\citenamefont {Tithof},
		\citenamefont {Suri}, \citenamefont {Pallantla}, \citenamefont {Grigoriev},\
		and\ \citenamefont {Schatz}}]{tithof_2017}%
	\BibitemOpen
	\bibfield  {author} {\bibinfo {author} {\bibfnamefont {J.}~\bibnamefont
			{Tithof}}, \bibinfo {author} {\bibfnamefont {B.}~\bibnamefont {Suri}},
		\bibinfo {author} {\bibfnamefont {R.~K.}\ \bibnamefont {Pallantla}}, \bibinfo
		{author} {\bibfnamefont {R.~O.}\ \bibnamefont {Grigoriev}}, \ and\ \bibinfo
		{author} {\bibfnamefont {M.~F.}\ \bibnamefont {Schatz}},\ }\href@noop {}
	{\bibfield  {journal} {\bibinfo  {journal} {J. Fluid Mech.}\ }\textbf
		{\bibinfo {volume} {828}},\ \bibinfo {pages} {837} (\bibinfo {year}
		{2017})}\BibitemShut {NoStop}%
	\bibitem [{\citenamefont {Egolf}\ \emph {et~al.}(2000)\citenamefont {Egolf},
		\citenamefont {Melnikov}, \citenamefont {Pesch},\ and\ \citenamefont
		{Ecke}}]{egolf_2002}%
	\BibitemOpen
	\bibfield  {author} {\bibinfo {author} {\bibfnamefont {D.~A.}\ \bibnamefont
			{Egolf}}, \bibinfo {author} {\bibfnamefont {I.~V.}\ \bibnamefont {Melnikov}},
		\bibinfo {author} {\bibfnamefont {W.}~\bibnamefont {Pesch}}, \ and\ \bibinfo
		{author} {\bibfnamefont {R.~E.}\ \bibnamefont {Ecke}},\ }\href@noop {}
	{\bibfield  {journal} {\bibinfo  {journal} {Nature}\ }\textbf {\bibinfo
			{volume} {404}},\ \bibinfo {pages} {733} (\bibinfo {year}
		{2000})}\BibitemShut {NoStop}%
	\bibitem [{\citenamefont {Karimi}\ and\ \citenamefont
		{Paul}(2012)}]{karimi_2012}%
	\BibitemOpen
	\bibfield  {author} {\bibinfo {author} {\bibfnamefont {A.}~\bibnamefont
			{Karimi}}\ and\ \bibinfo {author} {\bibfnamefont {M.~R.}\ \bibnamefont
			{Paul}},\ }\href {\doibase 10.1103/PhysRevE.85.046201} {\bibfield  {journal}
		{\bibinfo  {journal} {Phys. Rev. E}\ }\textbf {\bibinfo {volume} {85}},\
		\bibinfo {pages} {046201} (\bibinfo {year} {2012})}\BibitemShut {NoStop}%
	\bibitem [{\citenamefont {Suri}\ \emph {et~al.}(2018)\citenamefont {Suri},
		\citenamefont {Tithof}, \citenamefont {Grigoriev},\ and\ \citenamefont
		{Schatz}}]{suri_2018}%
	\BibitemOpen
	\bibfield  {author} {\bibinfo {author} {\bibfnamefont {B.}~\bibnamefont
			{Suri}}, \bibinfo {author} {\bibfnamefont {J.}~\bibnamefont {Tithof}},
		\bibinfo {author} {\bibfnamefont {R.~O.}\ \bibnamefont {Grigoriev}}, \ and\
		\bibinfo {author} {\bibfnamefont {M.~F.}\ \bibnamefont {Schatz}},\ }\href
	{\doibase 10.1103/PhysRevE.98.023105} {\bibfield  {journal} {\bibinfo
			{journal} {Phys. Rev. E}\ }\textbf {\bibinfo {volume} {98}},\ \bibinfo
		{pages} {023105} (\bibinfo {year} {2018})}\BibitemShut {NoStop}%
	\bibitem [{\citenamefont {Kazantsev}(1998)}]{kazantsev_1998}%
	\BibitemOpen
	\bibfield  {author} {\bibinfo {author} {\bibfnamefont {E.}~\bibnamefont
			{Kazantsev}},\ }\href@noop {} {\bibfield  {journal} {\bibinfo  {journal}
			{Nonlinear Proc. Geoph.}\ }\textbf {\bibinfo {volume} {5}},\ \bibinfo {pages}
		{193} (\bibinfo {year} {1998})}\BibitemShut {NoStop}%
	\bibitem [{\citenamefont {Kerswell}\ and\ \citenamefont
		{Tutty}(2007)}]{kerswell_2007}%
	\BibitemOpen
	\bibfield  {author} {\bibinfo {author} {\bibfnamefont {R.~R.}\ \bibnamefont
			{Kerswell}}\ and\ \bibinfo {author} {\bibfnamefont {O.~R.}\ \bibnamefont
			{Tutty}},\ }\href {\doibase 10.1017/S0022112007006301} {\bibfield  {journal}
		{\bibinfo  {journal} {J. Fluid Mech.}\ }\textbf {\bibinfo {volume} {584}},\
		\bibinfo {pages} {69–102} (\bibinfo {year} {2007})}\BibitemShut {NoStop}%
	\bibitem [{\citenamefont {Zoldi}\ and\ \citenamefont
		{Greenside}(1998)}]{zoldi_1998}%
	\BibitemOpen
	\bibfield  {author} {\bibinfo {author} {\bibfnamefont {S.~M.}\ \bibnamefont
			{Zoldi}}\ and\ \bibinfo {author} {\bibfnamefont {H.~S.}\ \bibnamefont
			{Greenside}},\ }\href {\doibase 10.1103/PhysRevE.57.R2511} {\bibfield
		{journal} {\bibinfo  {journal} {Phys. Rev. E}\ }\textbf {\bibinfo {volume}
			{57}},\ \bibinfo {pages} {R2511} (\bibinfo {year} {1998})}\BibitemShut
	{NoStop}%
	\bibitem [{\citenamefont {Zoldi}(1998)}]{zoldi_1998b}%
	\BibitemOpen
	\bibfield  {author} {\bibinfo {author} {\bibfnamefont {S.~M.}\ \bibnamefont
			{Zoldi}},\ }\href {\doibase 10.1103/PhysRevLett.81.3375} {\bibfield
		{journal} {\bibinfo  {journal} {Phys. Rev. Lett.}\ }\textbf {\bibinfo
			{volume} {81}},\ \bibinfo {pages} {3375} (\bibinfo {year}
		{1998})}\BibitemShut {NoStop}%
	\bibitem [{\citenamefont {Schneider}\ \emph {et~al.}(2007)\citenamefont
		{Schneider}, \citenamefont {Eckhardt},\ and\ \citenamefont
		{Vollmer}}]{schneider_2007a}%
	\BibitemOpen
	\bibfield  {author} {\bibinfo {author} {\bibfnamefont {T.~M.}\ \bibnamefont
			{Schneider}}, \bibinfo {author} {\bibfnamefont {B.}~\bibnamefont {Eckhardt}},
		\ and\ \bibinfo {author} {\bibfnamefont {J.}~\bibnamefont {Vollmer}},\ }\href
	{\doibase 10.1103/PhysRevE.75.066313} {\bibfield  {journal} {\bibinfo
			{journal} {Phys. Rev. E}\ }\textbf {\bibinfo {volume} {75}},\ \bibinfo
		{pages} {066313} (\bibinfo {year} {2007})}\BibitemShut {NoStop}%
	\bibitem [{\citenamefont {Farano}\ \emph {et~al.}(2018)\citenamefont {Farano},
		\citenamefont {Cherubini}, \citenamefont {Robinet}, \citenamefont
		{De~Palma},\ and\ \citenamefont {Schneider}}]{farano_2018a}%
	\BibitemOpen
	\bibfield  {author} {\bibinfo {author} {\bibfnamefont {M.}~\bibnamefont
			{Farano}}, \bibinfo {author} {\bibfnamefont {S.}~\bibnamefont {Cherubini}},
		\bibinfo {author} {\bibfnamefont {J.-C.}\ \bibnamefont {Robinet}}, \bibinfo
		{author} {\bibfnamefont {P.}~\bibnamefont {De~Palma}}, \ and\ \bibinfo
		{author} {\bibfnamefont {T.~M.}\ \bibnamefont {Schneider}},\ }\href {\doibase
		10.1017/jfm.2018.860} {\bibfield  {journal} {\bibinfo  {journal} {Journal of
				Fluid Mechanics}\ }\textbf {\bibinfo {volume} {858}},\ \bibinfo {pages} {R3}
		(\bibinfo {year} {2018})}\BibitemShut {NoStop}%
	\bibitem [{\citenamefont {L{\"u}thje}\ \emph {et~al.}(2001)\citenamefont
		{L{\"u}thje}, \citenamefont {Wolff},\ and\ \citenamefont
		{Pfister}}]{luthje_2001}%
	\BibitemOpen
	\bibfield  {author} {\bibinfo {author} {\bibfnamefont {O.}~\bibnamefont
			{L{\"u}thje}}, \bibinfo {author} {\bibfnamefont {S.}~\bibnamefont {Wolff}}, \
		and\ \bibinfo {author} {\bibfnamefont {G.}~\bibnamefont {Pfister}},\
	}\href@noop {} {\bibfield  {journal} {\bibinfo  {journal} {Phys. Rev. Lett.}\
		}\textbf {\bibinfo {volume} {86}},\ \bibinfo {pages} {1745} (\bibinfo {year}
		{2001})}\BibitemShut {NoStop}%
\end{thebibliography}
\end{document}